\documentstyle[aps,pre,amsmath,psfig,floats]{revtex}

\newcommand{\beq}{\begin{equation}}
\newcommand{\eeq}{\end{equation}}
\newcommand{\beqa}{\begin{eqnarray}}
\newcommand{\eeqa}{\end{eqnarray}}

\newcommand\ave[1]{\langle#1\rangle}
\newcommand\bave[1]{\big\langle#1\big\rangle}

\newcommand\dave[1]{\langle\hspace{-0.25em}\langle{#1}\rangle\hspace{-0.25em}\rangle}
\newcommand\bdave[1]{\big\langle\hspace{-0.32em}\big\langle{#1}\big\rangle\hspace{-0.32em}\big\rangle}

\renewcommand{\H}{{\cal H}}
\newcommand{\U}{{\cal U}}
\newcommand{\D}{{\cal D}}
\newcommand{\Z}{{\cal Z}}
\newcommand{\Sm}{{\cal S}}
\renewcommand{\P}{{\cal P}}
\renewcommand{\O}{{\cal O}}

\newcommand{\bhp}{{\boldsymbol{\hat{p}}}}
\newcommand{\bhq}{{\boldsymbol{\hat{q}}}}
\newcommand{\bp}{{\boldsymbol p}}
\newcommand{\bq}{{\boldsymbol q}}
\newcommand{\bxi}{{\boldsymbol\xi}}
\newcommand{\bz}{{\boldsymbol z}}

\newcommand{\bA}{{\boldsymbol A}}
\newcommand{\bB}{{\boldsymbol B}}
\newcommand{\bC}{{\boldsymbol C}}
\newcommand{\bU}{{\boldsymbol U}}
\newcommand{\bkappa}{{\boldsymbol K}}
\newcommand{\bLambda}{{\boldsymbol\Lambda}}
\newcommand{\bPsi}{{\boldsymbol\Psi}}

\begin{document}

 \wideabs{
% {
\title{Quantum thermodynamics of systems with anomalous dissipative
 coupling}
\author{Alessandro Cuccoli\cite{e-AC}, Andrea Fubini\cite{e-AF},
        Valerio Tognetti\cite{e-VT}}
\address{Dipartimento di Fisica dell'Universit\`a di Firenze
    and Istituto Nazionale di Fisica della Materia (INFM),
    \\ Largo E. Fermi~2, I-50125 Firenze, Italy}
\author{Ruggero Vaia\cite{e-RV}}
\address{Istituto di Elettronica Quantistica
    del Consiglio Nazionale delle Ricerche,
    via Panciatichi~56/30, I-50127 Firenze, Italy,
    \\ and Istituto Nazionale di Fisica della Materia (INFM)}
\date{\today}
\maketitle

\begin{abstract}
The standard {\em system-plus-reservoir} approach used in the study of
dissipative systems can be meaningfully generalized to a dissipative
coupling involving the momentum, instead of the coordinate: the
corresponding equation of motion differs from the Langevin equation, so
this is called {\em anomalous} dissipation. It occurs for systems where
such coupling can indeed be derived from the physical analysis of the
degrees of freedom which can be treated as a dissipation bath. Starting
from the influence functional corresponding to anomalous dissipation,
it is shown how to derive the effective classical potential that gives
the quantum thermal averages for the dissipative system in terms of
classical-like calculations; the generalization to many degrees of
freedom is given. The formalism is applied to a single particle in a
double-well and to the discrete $\phi^4$ model. At variance with the
standard case, the fluctuations of the coordinate are enhanced by
anomalous dissipative coupling.
\end{abstract}

%\vspace{10mm}
%\pacs{03.65.Yz, 05.30.-d, 05.40.-a, 64.60.Cn}
% 03.65.Yz Decoherence; open systems; quantum statistical methods
% 05.20.-y Statistical mechanics
% 05.30.-d Quantum statistical mechanics
% 05.40.-a Fluctuation phenomena, random processes, and Brownian motion
% 05.45.Yv Solitons
% 05.70.-a Thermodynamics
% 64.60.Cn Order-disorder transformations; statistical mechanics of model systems
% 65.40.Ba Heat capacity (crystalline solids)
% 65.90.+i Other topics in thermal properties of condensed matter

} %ends \wideabs
% \narrowtext

%%%%%%%%%%%%%%%%%%%%%%%%%%%%%%%%%%%%%%%%%%%%%%%%%%%%%%%%%%%%%%%%%%%%%%%%%%%%%
\section{Introduction}

The {\em system-plus-reservoir} (SPR)
model~\cite{Ullersma1966,Zwanzig1973,CaldeiraL1983,FordLO1988,Weiss1999}
is the most common and effective approach to the treatment of quantum
dissipation. This model realistically assumes the dissipation in a
system as due to the interaction with a reservoir (or bath, or
environment): anyone of the numerous degrees of freedom of the
reservoir is only weakly perturbed, so the reservoir is at thermal
equilibrium; moreover, in order to reproduce the general
phenomenological dynamics (quantum Langevin equation for the
coordinate) it is sufficient to assume a linear interaction and a bath
of harmonic oscillators~\cite{FordLO1988}, so that in the
imaginary-time path-integral formalism it is possible to integrate out
the bath variables and get a reduced description of the system in terms
of a bilocal {\em influence action}~\cite{Weiss1999}. Starting from
this framework, in previous papers we obtained the classical effective
potential for the calculation of thermal averages in a dissipative
quantum system~\cite{CRTV1997} and gave applications to the
$\phi^4$-chain model~\cite{CFTV1999} and to a 2D array of Josephson
junctions~\cite{CFTV2000}.

At variance with the standard approach of coupling the bath with the
system's coordinate, we consider in this paper the complementary
possibility of a coupling with the momentum, that Leggett called {\em
anomalous dissipation}~\cite{Leggett1984}. In this case the dynamics
can be reduced to a pseudo-Langevin equation where the dissipative term
contains the second derivative of the potential. This raises the
question whether one can tell the dissipation mechanism just from the
phenomenological behavior: we believe that one should have a physically
meaningful microscopic model for the bath rather than rely upon the
phenomenological counterpart. The case of anomalous dissipation occurs,
for instance, when the effect of the blackbody electromagnetic field on
a Josephson junction is considered~\cite{SolsZ1997}.

As for thermodynamics, we will show that the {\em anomalous influence
action} can be derived following the original
Feynman-Vernon~\cite{FeynmanV1963} idea, which, due to the
momentum-path dependence, can be pursued at the price of involving the
full phase-space path integral.

The main goal of this paper is in the treatment of this path integral
within the effective-potential formalism~\cite{CGTVV1995}, that reduces
the evaluation of quantum-dissipative thermal averages to much simpler
classical-like configuration integrals. This is accomplished also in
the case of many degrees of freedom, making quantitative calculations
feasible. In general, the inclusion of dissipation by coupling the
system's momenta with the environment results in higher quantum
fluctuations of the coordinates, while those of the momenta are
suppressed.

We introduce in Section~\ref{s.anomalous} the concept of anomalous
dissipation; it turns out that the influence action depends on the
momentum path and it is hence necessary to use the Hamiltonian path
integral. In Section~\ref{s.Veff} we treat it within the pure-quantum
self-consistent harmonic approximation~\cite{CGTVV1995}, deriving the
classical effective potential and the corresponding classical-like
expressions for quantum thermal averages; in the case of many degrees
of freedom the treatment is simplified by the additional `low-coupling'
approximation. Eventually, in section~\ref{s.applications} we show how
the framework works for the single particle in a double-well potential
and for the $\phi^4$-chain.

%================================================================
\section{From standard to anomalous dissipation}
\label{s.anomalous}

The {\em standard} SPR Hamiltonian has the form
\begin{equation}
 \hat\H = \frac {\hat p^2}{2m} + V(\hat q)
 + \frac12 \sum_{\ell} \Big[ \frac {\hat p_\ell^2}{m_\ell}
 + m_\ell\omega_\ell^2 (\hat q_\ell-\hat q)^2 \Big]~,
\label{e.sSPR}
\end{equation}
where $\hat{p}$ and $\hat{q}$ are the momentum and coordinate of the
`system' (here, one single particle in the one-dimensional potential
$V$), while $\hat{p}_\ell$ and $\hat{q}_\ell$ are those of the
$\ell$-th degree of freedom of the reservoir. It is quite a general
result~\cite{FordLO1988} that a suitable distribution of the bath
parameters $\{m_\ell,\omega_\ell\}$ (see Appendix~\ref{a.sSPR}) can
reproduce, when the bath variables are eliminated from the equations of
motion, the most general quantum Langevin equation namely,
\begin{equation}
 m\ddot{\hat q} + \int_{-\infty}^t dt'~\gamma(t{-}t')~\dot{\hat q}(t')
 + V'(\hat q) = \hat F(t)~.
\label{e.sLangevin}
\end{equation}
Since it does not explicitly contain the `microscopic' degrees of
freedom, the Langevin equation constitutes a macroscopic description of
the dissipative system with a clear phenomenological meaning, in the
sense that the memory function $\gamma(t)$ can be thought to be
determined experimentally. The thermodynamic density matrix for the
standard SPR model at the equilibrium temperature $T=\beta^{-1}$ has
the path-integral expression~\eqref{e.srho}, where dissipation is
described by the additional {\em influence action}
\begin{equation}
 S_{{}_{\rm{I}}}[q(u)] = \int_0^{\beta\hbar}\! \frac{du}{2\hbar}
 \int_0^{\beta\hbar}\! \frac{du'}{\beta\hbar}~k(u{-}u')~q(u)\,q(u')~,
\label{e.sinfluence}
\end{equation}
and the Matsubara transform of the kernel $k(u)$ is directly related
with the Laplace transform of the memory function as
$k_n=|\nu_n|~\tilde\gamma\big(|\nu_n|\big)$, where
$\nu_n=2\pi{n}/\beta\hbar$. For a harmonic potential one obtains for
the density matrix the exact result~\eqref{e.sSPR-HA}, so it appears
that standard dissipation quenches $\ave{\hat{q}^2}$ and rises
$\ave{\hat{p}^2}$. This breaks the canonical symmetry between the
coordinate and the momentum, and is obviously a consequence of the bath
being coupled to the coordinate in the standard SPR
model~(\ref{e.sSPR}). In the general case of a nonlinear interaction
$V(q)$ the evaluation of the path integral~(\ref{e.srho}) for the SPR
model becomes quite complicated, and was the subject of
Refs.~\cite{CRTV1997} and~\cite{CFTV1999}, where a classical effective
potential suitable to reduce the problem to classical-like expressions
was introduced as a generalization of the original approach of
Refs.~\cite{GT1985prl} and~\cite{FeynmanK1986}.
\medskip

The different case of {\em anomalous} dissipation occurs when the SPR
model is modified including the momentum in the place of the coordinate
in the interaction with the reservoir. The dependence upon the bath
coordinates and momenta is quadratic, so they can be exchanged at will
and the {\em anomalous} SPR Hamiltonian can be written
\begin{equation}
 \hat\H = \frac {\hat p^2}{2m} + V(\hat q)
 + \frac12 \sum_{\ell} \Big[ \frac {(\hat p_\ell-\hat p)^2}{m_\ell}
 + m_\ell\omega_\ell^2\,\hat q_\ell^2 \Big]~.
\label{e.aSPR}
\end{equation}
It is immediately seen that Eq.~(22) of Ref.~\onlinecite{SolsZ1997},
regarding a single Josephson junction interacting with the blackbody
electromagnetic field (in the dipole approximation), can be cast
exactly in this form, constituting a first physical example of the
relevance of treating anomalous dissipation.

Let us look at the dynamics of the system~\eqref{e.aSPR}: as in the
standard SPR case, one can derive the equations of motion and then
eliminate the bath variables. The result is a {\em pseudo}-Langevin
equation,
\begin{equation}
 m\ddot{\hat q} + m\int_{-\infty}^t dt'
 ~\eta(t{-}t')~\partial_{t'}V'\big(\hat q(t')\big)
 + V'(\hat q) = \hat F(t)~.
\label{e.aLangevin}
\end{equation}
To interpret Eq.~(\ref{e.aLangevin}), observe that in the classical
limit one has
$\partial_{t'}V'\big(q(t')\big)=V''\big(q(t')\big)~\dot{q}(t')$, so
that in the neighborhood of the stable minimum, where
$V''(q)\sim{m\omega^2}$ tends to a positive constant, it reduces to the
standard form of the Langevin equation~(\ref{e.sLangevin}); for a
harmonic potential this always occurs. The physical difference is that
the full damping function $\gamma(t)=m^2\omega^2\eta(t)$ depends on the
system's mass and potential, not barely on the reservoir's
characteristics, so that the frictional force depends on how the
system's motion arises, which prevents from a simple phenomenological
interpretation~\cite{FordLO1988}. On the same footing, the concept of
anomalous dissipation was considered by Caldeira and Leggett (see
Appendix C of Ref.~\cite{CaldeiraL1983}) and further analyzed by
Leggett~\cite{Leggett1984}. Their purpose was to establish that in the
context of tunneling problems the standard SPR model~(\ref{e.sSPR})
(with a possible coupling to a nonlinear function of the coordinate) is
the most general one to be considered in order to infer the effects of
damping from a knowledge of the phenomenological quasiclassical
dissipative equation~\eqref{e.sLangevin}; therefore, they rule out the
anomalous case as it can give a negative friction coefficient ($V''<0$)
in the pseudo-Langevin equation over much of the tunneling region,
which would lead to qualitatively different results. On the other hand,
since the pseudo-Langevin equation reduces to the Langevin one in the
dynamical asymptotic region (i.e., close to the equilibrium
configuration), a purely phenomenological approach to dissipation based
on the dynamical linear response cannot distinguish whether the
underlying dissipative mechanism is of the standard or of the anomalous
type, while this distinction plays a fundamental role in the quantum
statistical mechanics. However, rather than being a schematization for
a phenomenological dissipative behavior, the mechanism can happen to be
determined from a physically sound microscopic model which can have the
form (\ref{e.aSPR}): once such microscopic SPR Hamiltonian is known,
there is no point in trying to reduce it to a phenomenological
description. Although examples can be given of how anomalous
dissipation can arise just reinterpreting the canonical variables in a
standard SPR system (e.g., a linear dissipative system as a simple RLC
circuit~\cite{Leggett1984}, using the canonical symmetry between
magnetic flux and charge), it is more significant that the anomalous
dissipative mechanism is obtained on a {\em microscopic} basis for
Josephson junctions, when quantum electromagnetic fluctuations are
considered in the dipole approximation~\cite{SolsZ1997}.

In order to get a first insight about the dissipation effect onto the
thermodynamics, let us consider the harmonic oscillator. In
Appendix~\ref{a.aSPR} the analytic solution is easily obtained using
the canonical symmetry between coordinate and momentum. At variance
with the standard case, it appears that anomalous dissipation quenches
the fluctuation of the momentum, Eq.~\eqref{e.alambdaT}, and enhances
those of the coordinate, Eq~\eqref{e.aalphaT}, as it is expected since
the environment now performs a `measurement' of the momentum. This kind
of behavior is expected also when the system is subjected to a
nonlinear potential~\cite{Leggett1984}, but in this case the evaluation
of quantum-dissipative thermal averages is by no means straightforward,
and one has to resort to some approximation scheme. The purpose of the
next Section is indeed to derive the effective classical potential and
the related prescriptions in order to evaluate the system's
thermodynamic averages for the anomalous SPR model~(\ref{e.aSPR}).

In writing the path integral for the density matrix corresponding to
the Hamiltonian~(\ref{e.aSPR}), it appears that in order to integrate
out the bath variables one must resort to the full phase-space
formulation for the system, so that the momentum path $p(u)$ enters the
interaction with the bath. Indeed, the {\em anomalous} influence action
turns out to be (see Appendix~\ref{a.SI})
\begin{equation}
 S_{{}_{\rm{I}}}[p(u)] = \int_0^{\beta\hbar}\! \frac{du}{2\hbar}
 \int_0^{\beta\hbar}\! \frac{du'}{\beta\hbar}~\kappa(u{-}u')~p(u)\,p(u')~,
\label{e.aSI}
\end{equation}
and the kernel $\kappa(u)=\sum_n\kappa_ne^{i\nu_nu}$ is now directly
related with the Laplace transform of the memory function $\eta(t)$ as
\begin{equation}
 \kappa_n = |\nu_n|~\tilde\eta\big(|\nu_n|\big)~.
\label{e.akneta}
\end{equation}
Note that $\kappa(u)$ is periodic and, since $\kappa_n=\kappa_{-n}$,
also even, so one has the properties
$\kappa(u)=\kappa(-u)=\kappa(u+\beta\hbar)$; in addition $\kappa_0=0$,
meaning that the influence action does not include a local component,
which trivially renormalizes the mass.

For the phase-space path integral it is convenient to use the symmetric
(Weyl) ordering prescription~\cite{Berezin1980,CTVV1992ham}. The Weyl
symbol for the density operator can therefore be written as
\begin{equation}
 \rho(p,q)=\int \D [p_u,q_u]
 ~e^{-S[p_u,q_u]-S_{{}_{\rm{I}}}[p_u]} ~,
\label{e.arho}
\end{equation}
where for compactness the arguments are indicated as subscripts, and
the nondissipative part of the system's action reads
\begin{eqnarray}
 & & S[p_u,q_u] =
 \int_0^{\beta\hbar}\!\frac{du}{\hbar}\,
 \Big[\, \frac i2 \big(q_u\dot p_u{-}p_u\dot q_u\big)
 +\frac {p_u^2}{2m}+V\big(q_u\big) \Big]
\nonumber\\ & & \hspace{5mm}
 +\frac i{2\hbar}( q_0p_\beta {-} p_0q_\beta )
 +\frac i\hbar\big[ (q_\beta{-}q_0)\,p - (p_\beta{-}p_0)\,q \big] ~.
\label{e.Spq}
\end{eqnarray}
Note that the arguments $p$ and $q$ do appear explicitly in the action,
while there are no constraints upon the paths; in particular, the end
points ($p_0$, $p_\beta$, $q_0$, $q_\beta$) are also integrated
over~\cite{Berezin1980}. The thermal average of an observable $\hat\O$
can be written in terms of its Weyl symbol,
\begin{equation}
 \O(p,q)=\int dx~e^{-i\,p\,x/\hbar} \Big\langle q{+}\frac x2 \Big|
 \,\hat\O\, \Big| q{-}\frac x2 \Big\rangle ~,
\end{equation}
as
\begin{equation}
 \ave{\hat\O} = \frac1\Z \int \frac {dp\,dq}{2\pi\hbar}
 ~\rho(p,q)~\O(p,q)~.
\label{e.aveOdef}
\end{equation}

%================================================================
\section{Effective potential}
\label{s.Veff}

\subsection{One degree of freedom}

The strategy~\cite{CGTVV1995} to approximate the density~(\ref{e.arho})
is to use the quadratic trial action $S_0[p_u,q_u;\bar{q}]$ expressed
as Eq.~(\ref{e.Spq}) where $V(q_u)$ is replaced by the trial
`potential'
\begin{equation}
 V_0(q_u;\bar{q})=w(\bar{q})
 +\frac12\,m\omega^2\,(\bar{q})\,(q_u-\bar{q})^2~,
\label{e.V0}
\end{equation}
which is indeed a functional through its dependence on the path's
average point
\begin{equation}
 \bar{q}[q_u] = \int_0^{\beta\hbar} \frac{du}{\beta\hbar}\,~q_u~,
\end{equation}
and to optimize its parameters $w(\bar{q})$ and $\omega^2(\bar{q})$.

The reduced density
\begin{eqnarray}
 \bar\rho_0(p,q;\bar{q}) &=& \int \D [p_u,q_u]
 ~{\textstyle \delta\big( \bar{q}-\int_0^{\beta\hbar}\!
 \frac{du}{\beta\hbar}\,q_u \big) }
\nonumber\\
 & & \hspace{10mm}\times
 ~ e^{-S_0[p_u,q_u;\bar{q}]-S_{{}_{\rm{I}}}[p_u]}
\label{e.barrho}
\end{eqnarray}
collects the contribution of classes of paths that share the same
average point $\bar{q}$, so the path integral~(\ref{e.arho}) for the
trial action $S_0$ can be cast into the form
\begin{equation}
 \rho_0(p,q)=\int d\bar{q} ~\bar\rho_0(p,q;\bar{q}) ~.
\label{e.arho1}
\end{equation}
What we are going to do is to optimize the reduced density $\bar\rho$.
This is an important point, that justifies the accuracy of the
approximation: we choose the best approximating Gaussian distribution
for the {\em separate} contribution of each class of paths (labeled by
$\bar{q}$), which describes the purely quantum-dissipative contribution
to the fluctuations around any `classical' configuration $\bar{q}$. In
other words, the classical fluctuations are accounted for exactly.

The reduced density can be evaluated analytically: by including the
constraint for the variable $\bar{q}$ in the action through the Fourier
representation of the Dirac delta function it is possible to get an
integral expression in terms of the dissipative harmonic oscillator
density matrix. In Appendix~\ref{a.1df.rho0p} we report the actual
calculations that lead to the final result, which is a Gaussian
distribution in $(p,q)$:
\begin{equation}
 \bar\rho_0(p,q;\bar{q})=\sqrt{\frac {2\pi m}\beta}~
 ~e^{-\beta V_{\rm{eff}}(\bar{q})}
 ~\frac {e^{-p^2/2\lambda}} {\sqrt{2\pi\lambda}}
 ~\frac {e^{-(q-\bar{q})^2/2\alpha}} {\sqrt{2\pi\alpha}}~,
\label{e.barrho1}
\end{equation}
where the effective potential reads
\begin{equation}
 V_{\rm{eff}}(\bar{q}) = w(\bar{q}) + \frac1\beta\, \mu(\bar{q}) ~,
\label{e.Veff1}
\end{equation}
with
\begin{equation}
 \mu(\bar{q}) = \sum_{n=1}^\infty
 \ln \frac{\nu_n^2+(1+m \kappa_n)\,\omega^2(\bar{q})}{\nu_n^2}~,
\label{e.mu}
\end{equation}
and the relevant variances are
\begin{eqnarray}
 \lambda(\bar{q}) &=&
 \frac {m}\beta \sum_{n=-\infty}^\infty
 \frac {\omega^2(\bar{q})}{\nu_n^2+(1+m \kappa_n)\,\omega^2(\bar{q})} ~.
\label{e.lambda}
\\
 \alpha(\bar{q}) &=&
 \frac 2{m\beta} \sum_{n=1}^\infty
 \frac {1+m \kappa_n}{\nu_n^2+(1+m \kappa_n)\,\omega^2(\bar{q})} ~;
\label{e.alpha}
\end{eqnarray}
apart from the dependence on the variational parameter
$\omega^2(\bar{q})$, these variances have the harmonic-oscillator
form~(\ref{e.alambdaT}) and~(\ref{e.aalphaT}) with the important
difference that $\alpha$ lacks the $n{=}0$ term, i.e., its classical
contribution.

We introduce henceforth the coordinate fluctuation variable
$\xi=q-\bar{q}$ in the place of $q$ and the following double-bracket
notation for the Gaussian averages over the variables $p$ and $\xi$
defined by $\bar{\rho}_0$:
\begin{equation}
 \bdave{p^2} = \lambda(\bar{q}) ~, ~~~~~
 \bdave{\xi^2} = \alpha(\bar{q})~.
\end{equation}
The physical interpretation of the above formulas becomes transparent
if one expresses the resulting approximation for the thermal
average~(\ref{e.aveOdef}) using Eqs.~(\ref{e.arho1})
and~(\ref{e.barrho1}), namely
\begin{equation}
 \ave{\hat{\O}} = \frac1\Z~\sqrt{\frac {m}{2\pi\hbar^2\beta}}~
 \int d\bar{q}~e^{-\beta\,V_{\rm{eff}}(\bar{q})}
 ~\bdave{O(p,\bar{q}+\xi)}~;
\label{e.aveO }
\end{equation}
the average over $\xi$ accounts for the nonclassical fluctuations of
the coordinate, while the classical part is accounted for exactly by
the classical-like expression for the thermodynamic average; the
average over $p$ keeps into account the full fluctuation, since its
classical contribution is exactly Gaussian. The detailed discussion
made in Ref.~\cite{CGTVV1995} applies in this case as well, provided
one keeps into account that the contribution of the dissipative
nonlocal action is also accounted for by the variances $\lambda$ and
$\alpha$.

The last point concerns the optimization of the parameters $w(\bar{q})$
and $\omega^2(\bar{q})$. This can be performed imposing that the trial
and the true potential (and their derivatives up to the second one)
have the same $\bar\rho_0$-average~\cite{CTVV1992ham,CGTVV1995}:
\begin{eqnarray}
 \bdave{V(\bar{q}{+}\xi)} &=&
 w(\bar{q}) + \frac12 m\,\omega^2(\bar{q})\, \alpha(\bar{q})\,,
\label{e.pqscha0}
\\
 \bdave{V''(\bar{q}{+}\xi)} &=& m\,\omega^2(\bar{q}) ~,
\label{e.pqscha2}
\end{eqnarray}
while the condition for the first derivative is trivial, since a linear
term $\sim(q-\bar{q})$ in $V_0$ does not contribute to the action by
the very definition of the average point. Eq.~(\ref{e.pqscha2}) is a
self-consistent equation that identifies the parameter
$\omega^2(\bar{q})$, while the determination of $w(\bar{q})$ in
Eq.~(\ref{e.pqscha0}) permits to rewrite the effective potential as
\begin{equation}
 V_{\rm{eff}}(\bar{q}) = \bdave{V(\bar{q}+\xi)}
  - \frac12 m\,\omega^2(\bar{q})~ \alpha(\bar{q})
  + \frac1\beta\, \mu(\bar{q})~.
\label{e.Veff2}
\end{equation}

The differential operator
\begin{equation}
 \Delta=\frac12\, %\big[ \lambda(\bar{q})~\partial_p^2 +
 \alpha(\bar{q})~\partial_{\bar{q}}^2 ~,
 %\big]~,
\label{e.Delta}
\end{equation}
understood as not operating onto its coefficient, is such that
$\Delta{F}(\bar{q})=\bdave{F(\bar{q}{+}\xi)}$, and allows us to rewrite
the effective potential in a more compact way:
\begin{equation}
 V_{\rm{eff}}(\bar{q}) = (1-\Delta)\,e^\Delta\,V(\bar{q})
 + \frac1\beta\, \mu(\bar{q})~.
\label{e.VeffDelta}
\end{equation}
It appears that the first term gives corrections of order $\alpha^2$ to
the potential, while first order renormalizations arise from the last
term.

%================================================================
\subsection{Many degrees of freedom}
\label{s.manydof}

Let us now consider a general system with $N$ degrees of freedom, i.e.,
canonical coordinate and momentum operators
$\bhq=\{\hat{q}_i\}_{i=1,...,N}$ and $\bhp=\{\hat{p}_i\}_{i=1,...,N}$,
with the commutation relations $[\hat{q}_i,\hat{p}_j]=i\,\delta_{ij}$
(we set $\hbar=1$ in this subsection), and described by a Hamiltonian
with a quadratic kinetic energy and a nonlinear potential term,
\begin{equation}
 \hat{\cal{H}} = \frac12\, {}^{\rm{t}}\!\bhp\, \bA^2 \bhp + V(\bhq)~.
 \label{e.mdf.hatH}
\end{equation}
The real matrix $\bA^2=\{A_{ij}^2 \}$ is symmetric,
${}^{\rm{t}}\!\bA=\bA$, and positive definite. The influence
action for anomalous dissipation takes the form \begin{equation}
 S_{{}_{\rm{I}}}[\bp(u)] = \int_0^{\beta}\! \frac{du}{2}
 \int_0^{\beta}\! \frac{du'}{\beta}~
 {}^{\rm{t}}\!\bp(u)~\bkappa(u{-}u')~\bp(u')~,
\label{e.mdf.SI}
\end{equation}
where the kernel matrix $\bkappa(u)=\{\kappa_{ij}(u)\}$ is a real
symmetric matrix that replaces the scalar kernel $\kappa(u)$ of the
single-particle case; as a function of $u$ it keeps its symmetry and
periodicity, $\bkappa(u)=\bkappa(-u)=\bkappa(\beta{-}u)$, and satisfies
$\int_0^\beta{du}\,\bkappa(u)=0$. For instance, in the case of $N$
independent identical baths coupled to each momentum $\hat{p}_i$ one
simply has a diagonal kernel, $\kappa_{ij}(u)=\delta_{ij}\,\kappa(u)$;
this case will be considered for the application shown in
Section~\ref{ss.phi4}.

The Weyl symbol for system's density matrix is expressed as an
$N$-dimensional integral, \begin{equation}
 \rho(\bp,\bq) = \int d\bar\bq~ \bar\rho(\bp,\bq;\bar\bq) ~,
\end{equation}
in terms of the reduced density collecting the contributions of paths
sharing the average configuration $\bar\bq$,
\begin{eqnarray}
 \bar\rho(\bp,\bq;\bar\bq) &=& \int\D[\bp_u,\bq_u]
 ~{\textstyle \delta\big( \bar\bq-\int_0^{\beta}
 \frac{du}{\beta}\,\bq_u \big) }
\nonumber\\
 & & \hspace{5mm}\times
 ~ e^{-\Sm[\bp_u,\bq_u;\bp,\bq]
 -S[\bp_u,\bq_u;\bar\bq]-S_{{}_{\rm{I}}}[\bp_u]}~;
\label{e.mdf.brho}
\end{eqnarray}
here the action is separated in a part containing the external
variables $(\bp,\bq)$,
\begin{eqnarray}
 & & \Sm[\bp_u,\bq_u;\bp,\bq] = \int_0^{\beta}\!du\,
 \Big[\, \frac i2 \big({}^{\rm{t}}\!\bq_u\dot\bp_u{-}
 {}^{\rm{t}}\!\bp_u\dot\bq_u\big) \Big]
\nonumber\\
 & & \hspace{3mm}
 +\frac i2 ( {}^{\rm{t}}\!\bq_0\bp_\beta {-} {}^{\rm{t}}\!\bp_0\bq_\beta )
 +i \big[ {}^{\rm{t}}\!(\bq_\beta{-}\bq_0)\,\bp
 - {}^{\rm{t}}\!(\bp_\beta{-}\bp_0)\,\bq \big] ,
\label{e.mdf.SM}
\end{eqnarray}
and the part containing the system's Hamiltonian,
\begin{equation}
 S[\bp_u,\bq_u;\bar\bq] = \int_0^{\beta}\!du\,
 \Big[\, \frac12\, {}^{\rm{t}}\!\bp_u \bA^2 \bp_u + V\big(\bq_u\big)
 \Big]~.
\label{e.mdf.S}
\end{equation}

In order to evaluate the effective potential, the trial action $S_0$ is
defined as $S$ by replacing $V(\bq_u)$ by
\begin{equation}
 V_0(\bq_u;\bar\bq) = w(\bar\bq) + \frac12~
 {}^{\rm{t}}(\bq_u{-}\bar\bq) \bB^2(\bar\bq) (\bq_u{-}\bar\bq)~,
\label{e.mdf.V0}
\end{equation}
with a real symmetric matrix $\bB^2(\bar\bq)$. The calculation of the
corresponding $\bar\rho_0$, reported in Appendix~\ref{a.mdf.rho0p}, is
not a trivial extension of that for one degree of freedom, but can take
advantage of the results obtained in Ref.~\cite{CFTV1999} for the
standard SPR model.

The simplest way for writing the final result is to give the expression
of the thermal average of a generic observable $\hat\O$ in terms of its
Weyl symbol $\O(\bp,\bq)$. This fundamental formula approximates
quantum averages by means of a classical-like expression with the
effective potential $V_{\rm{eff}}$,
\begin{equation}
 \bave{\hat\O}=
 \frac1\Z \bigg(\frac1{2\pi\beta}\bigg)^{\!\!\frac N2}\frac1{\det\bA}
 \int d\bar\bq ~e^{-\beta V_{\rm{eff}}(\bar\bq)}
 \bdave{\O(\bp,\bar\bq{+}\bxi)} ~,
\label{e.mdf.aveO}
\end{equation}
where $\dave{\,\cdot\,}$ is the Gaussian average over the
variables $\bp$ and $\bxi$ determined by $\bar\rho_0$, as reported
in Eq.~(\ref{e.mdf.brho0}), and can be uniquely defined through
its moments \begin{eqnarray}
 \bdave{\bxi\,{}^{\rm{t}}\!\bxi}=\bC(\bar\bq) &=&
 \frac2\beta\sum_{n=1}^\infty
 \bB^{-1}\frac {\bPsi_n}{\nu_n^2+\bPsi_n}\,\bB^{-1} ~,
\label{e.Cij}\\
 \bdave{\bp\,{}^{\rm{t}}\!\bp}=\bLambda(\bar\bq) &=&
 \frac1\beta\,\sum_{n=-\infty}^\infty
 \bB\,\frac1{\nu_n^2+\bPsi_n}\,\bB ~,
\label{e.Lambdaij}
\end{eqnarray}
whose components are the {\em renormalization coefficients} and
\begin{equation}
 \bPsi_n(\bar\bq) = \bB\,\big(\bA^2 + \bkappa_n\big)\,\bB ~.
\end{equation}
The effective potential reads
\begin{equation}
 V_{\rm{eff}}(\bar\bq) = w(\bar\bq) + \frac 1\beta~ \mu(\bar\bq) ~,
\label{e.mdf.Veff}
\end{equation}
with
\begin{equation}
 \mu(\bar\bq) = \sum\limits_{n=1}^{\infty}\,
 \ln\,\frac{\det(\nu_n^2+\bPsi_n)}{\nu_n^{2N}}~.
\label{e.mdf.mu}
\end{equation}
To determine the parameters $w$ and $\bB(\bar\bq)$ we require, as
in Eqs.~(\ref{e.pqscha0}) and~(\ref{e.pqscha2}), that the
parameters of the trial action are such to match the
$\rho_0$-averages of the original and the trial potential, and the
same for their second derivatives:
\begin{eqnarray}
 \bdave{V(\bar\bq+\bxi)} &=&
 w(\bar\bq)+{\textstyle\frac12}\,{\rm{Tr}}\,
 \big[\bB^2(\bar\bq)\,\bC(\bar\bq)\big]  ~,
\label{e.w}
\\
 \bdave{\partial_{q_i}\partial_{q_j}V(\bar\bq+\bxi)}
 &=& B^2_{ij}(\bar\bq) ~.
\label{e.Bij}
\end{eqnarray}
The latter equation together with Eq.~(\ref{e.Cij}) self-consis\-tently
determines the solution for the matrices $\bB(\bar\bq)$ and
$\bC(\bar\bq)$. Its matrix character makes it useful to introduce the
`low-coupling' approximation (LCA), in the very same way of
Ref.~\cite{CFTV1999}, so it is sufficient to write here the final
results. If $\bar\bq_0$ is the configuration that minimizes
$V_{\rm{eff}}(\bar\bq)$, one has the LCA effective potential
\begin{equation}
 V_{\rm{eff}}(\bar\bq) =
 e^\Delta\,V(\bar\bq) -\Delta e^\Delta V(\bar\bq_0) + \beta^{-1}\,\mu ~,
\end{equation}
where $\mu=\mu(\bar\bq_0)$ and the operator
\begin{equation}
 \Delta =
 \frac12 \sum_{ij}C_{ij}~\partial_{\bar{q}_i}\partial_{\bar{q}_j}~,
\label{e.mdf.Delta}
\end{equation}
with $\bC\equiv\bC(\bar\bq_0)$, is such that
$e^\Delta\,V(\bar\bq)=\bdave{V(\bar\bq{+}\bxi)}$ (within the LCA).
Therefore Eqs.~(\ref{e.Cij}) and~(\ref{e.Bij}) have to be solved only
for the minimum configuration, with a great simplification. In the
simplest case of translation invariance an orthogonal Fourier
transformation $\bU=\big\{U_{ki}\big\}$ diagonalizes all matrices,
\begin{eqnarray}
 m^{-1}_k~\delta_{kk'}&=&{\sum}_{ij}~U_{ki}\,U_{k'j}\,A^2_{ij} ~,
\label{e.Ak}
\\
 m_k\,\omega^2_k~\delta_{kk'}&=&{\sum}_{ij}~U_{ki}\,U_{k'j}\,B^2_{ij} ~,
\label{e.Bk}
\\
 \kappa_{n,k}~\delta_{kk'}&=&{\sum}_{ij}~U_{ki}\,U_{k'j}\,\kappa_{n,ij} ~,
\label{e.Kk}
\end{eqnarray}
so that
\begin{equation}
 \mu = \sum_k\,\sum_{n=1}^\infty
 \ln\frac{\nu_n^2+(1+m_k\kappa_{n,k})\,\omega^2_k}{\nu_n^2} ~;
\end{equation}
the renormalization coefficients of Eqs.~(\ref{e.Cij})
and~(\ref{e.Lambdaij}) become
\begin{equation}
 \Lambda_{ij} = {\sum}_{k}~U_{ki}\,U_{kj}\, \lambda_k~,~~~~
       C_{ij} = {\sum}_{k}~U_{ki}\,U_{kj}\, \alpha_k~,
\label{e.alphakLCA}
\end{equation}
where the variances of the $k$-th mode,
\begin{eqnarray}
 \lambda_k &=& \frac{m_k}\beta \sum_{n=-\infty}^\infty
 \frac{\omega^2_k}{\nu_n^2+(1+m_k\kappa_{n,k})\,\omega^2_k} ~,
\label{e.LambdakLCA}
\\
 \alpha_k &=& \frac{2}{\beta m_k}\, \sum_{n=1}^\infty
 \frac{1+m_k\kappa_{n,k}}{\nu_n^2+(1+m_k\kappa_{n,k})\,\omega^2_k} ~,
\label{e.CkLCA}
\end{eqnarray}
generalize those of Eqs.~(\ref{e.lambda}) and (\ref{e.alpha}); for
instance, the `on-site' renormalization coefficient $D\equiv{C_{ii}}$
can be simply expressed as $D=\bdave{\xi_i^2}=N^{-1}\sum_k\alpha_k$\,.
The partition function and thermal averages are to be evaluated by
means of Eq.~(\ref{e.mdf.aveO}), where, of course, the effective
potential and the double-bracket average are to be understood as the
LCA ones. Note that, since the LCA $\Lambda$'s do not depend on
$\bar\bq$, averages of observables involving only momenta are trivially
evaluated as $\bave{\O(\hat\bp)}=\bdave{\O(\bp)}$.

%================================================================
\section{Applications}
\label{s.applications}

\subsection{Drude response model}

In the case of {\it Ohmic dissipation} the memory is Markovian,
$\eta(t)=\eta\,\delta(t-0)$, corresponding to
$\tilde\eta(z)=\eta={\rm{const}}$ and then to $\kappa_n\propto|n|$, so
one can see that the coordinate fluctuation $\alpha(\bar{q})$,
Eq.~(\ref{e.alpha}), is divergent, as well as $\mu(\bar{q})$,
Eq.~(\ref{e.mu}). This is due to the unphysical assumption of a
vanishing response time from the dissipation bath; in the standard SPR
model such divergence~\cite{CaldeiraL1983,Weiss1999,CRTV1997,CFTV1999}
only affects the momentum fluctuation, leaving the possibility to
meaningfully evaluate thermal averages of coordinate-dependent
quantities~\cite{CRTV1997}. As when considering momentum-dependent
quantities~\cite{CFTV1999} in the standard case, in the anomalous SPR
model we are therefore forced to take into account the finite bath's
response time.

The Drude model of an exponentially decaying memory,
\begin{equation}
 \eta(t) = \Theta(t)\,\eta\,\omega_{{}_{\rm{D}}}
 e^{-\omega_{{}_{\rm{D}}}t}
~,~~~~
 \tilde\eta(z) =
 \frac{\eta\,\omega_{{}_{\rm{D}}}}{\omega_{{}_{\rm{D}}}+z}
\label{e.Drude}
\end{equation}
is the simplest one which is physically meaningful: the constant $\eta$
characterizes the strength of the coupling with the dissipation bath,
while the Drude frequency $\omega_{{}_{\rm{D}}}$ characterizes its
bandwidth ($\omega_{{}_{\rm{D}}}^{-1}$ is the response time); the Ohmic
limit occurs for $\omega_{{}_{\rm{D}}}\to{\infty}$.

%The constant $m\eta$ has the dimensions of a time and
%from Eq.~(\ref{e.akneta})
%\begin{equation}
% m\,k_n=|\nu_n|\,m\tilde\eta(|\nu_n|)
%\end{equation}

\subsection{Single particle in the double-well}
\label{ss.dwell}

Given a potential $V(q)$, it is convenient to provide a dimensionless
formulation by identifying characteristic energy and length scales,
$\epsilon$ and  $\sigma$. The first one could be a barrier height or a
well depth, the latter is such that the change of $V$ over its scale is
comparable to $\epsilon$. The dimensionless coordinate is
$\hat{x}=\hat{q}/\sigma$ and the dimensionless potential $v(x)$ is such
that $V(\sigma{x})=\epsilon\,v(x)$. If $x_{\rm{m}}$ is the absolute
minimum of $v(x)$ and $v''\equiv{v}''(x_{\rm{m}})$, the harmonic
approximation (HA) of the system is characterized by the frequency
$\omega_0=\sqrt{\epsilon\,v''/m\sigma^2}$; the ratio between the HA
quantum energy level splitting $\hbar\omega_0$ and the overall energy
scale $\epsilon$ defines the dimensionless coupling parameter
\begin{equation}
 g=\frac{\hbar\omega_0}{\epsilon}
 =\sqrt{\frac{\hbar^2 v''}{m\epsilon\sigma^2}}~.
\label{e.g}
\end{equation}
Therefore the system is weakly (strongly) `quantum' when $g$ is small
(large) compared to 1. Defining the dimensionless momentum
$\hat{p}_x=\sigma\hat{p}/\hbar$, such that $[\hat{x},\hat{p}_x]=i$, the
nondissipative part of the Hamiltonian (\ref{e.aSPR}) reads
\begin{equation}
 \frac{\hat\H_{{}_{\rm{S}}}}\epsilon =
 \frac{g^2}{v''} \, \frac {{\hat p}_x^2}2 + v(\hat{x}) ~.
\end{equation}

In what follows energies are given in units of $\epsilon$, lengths in
units of $\sigma$, frequencies in units of $\omega_0$, and so on; the
dimensionless temperature is $t=1/(\epsilon\beta)$.

As for the dissipation part, to make contact with the phenomenological
behavior, note that the asymptotic damping function from the classical
limit of the pseudo-Langevin equation (\ref{e.aLangevin}) is
$V''\tilde\eta(z)=m\omega_0^2\tilde\eta(z)$ and has the dimension of a
frequency [it is indeed the counterpart of $\tilde\gamma(z)/m$ of the
standard SPR model], so for the Drude form (\ref{e.Drude}) we will deal
with the corresponding dimensionless input parameters
\begin{equation}
 \hat\eta=m\omega_0\eta
~,~~~~~~~~
 \hat\omega_{{}_{\rm{D}}}=\omega_{{}_{\rm{D}}}/\omega_0 ~.
\end{equation}
The dimensionless term $m\kappa_n$ reads then
\begin{equation}
 m\kappa_n = \hat\eta\,\hat\omega_{{}_{\rm{D}}}
 \frac{\pi n}{\pi n+f_{{}_{\rm{D}}}} \equiv a_n ~,
\label{e.an}
\end{equation}
where $f_{{}_{\rm{D}}}=\beta\hbar\omega_{{}_{\rm{D}}}/2=
(g/2t)\hat\omega_{{}_{\rm{D}}}$.

Using the dimensionless parameter
\begin{equation}
 f(x) = \frac{\beta\hbar\omega(x)}2=
 \frac{g}{2t}\frac{\omega(x)}{\omega_0}
\end{equation}
in the place of $\omega(x)$, from the definitions~(\ref{e.mu}),
(\ref{e.lambda}), and (\ref{e.alpha}) we obtain
\begin{eqnarray}
 \mu(x)&=&\sum_{n=1}^{\infty} \ln\,
 \frac{(\pi n)^2+(1+a_n)\,f^2(x)}{(\pi n)^2}~,
\label{e.dwmu}
\\
 \lambda(x) &=& \frac{v''t}{g^2}\,\sum\limits_{n=-\infty}^{\infty}
 \frac{f^2(x)}{(\pi n)^2+(1+a_n)\,f^2(x)}~,
\label{e.dwlambda}
\\
 \alpha(x) &=& \frac{g^2}{2v''\,t}\,\sum\limits_{n=1}^{\infty}
 \frac{1+a_n}{(\pi n)^2+(1+a_n)\,f^2(x)}~,
\label{e.dwalpha}
\end{eqnarray}
where $\lambda=\bdave{p_x^2}$.

The example we choose here is the double-well potential,
\begin{equation}
 v(x)=(1-x^2)^2 ~,
\end{equation}
which has two degenerate symmetric minima in $x_{\rm{m}}=\pm{1}$ with
$v''=8$. From Eqs.~(\ref{e.VeffDelta}) and~(\ref{e.pqscha2}) we obtain
\begin{eqnarray}
 v_{\rm{eff}}(x)&=&(1-x^2)^2-3\,\alpha^2(x)+t\,\mu(x)~,
\\
 f^2(x) &=& \frac{g^2}{8\,t^2}\,[3\,x^2+3\,\alpha(x)-1]~.
\label{e.dwf2}
\end{eqnarray}
Eq.~(\ref{e.dwf2}) has to be solved self-consistently with
Eq.~(\ref{e.dwalpha}): we did this numerically; exact reference data
can be obtained only for the nondissipative limit $\hat\eta=0$, e.g.,
by numerically solving the Schr\"odinger equation.

\begin{figure}[hbt]
\centerline{\psfig{bbllx=13mm,bblly=33mm,bburx=186mm,bbury=256mm,%
 figure=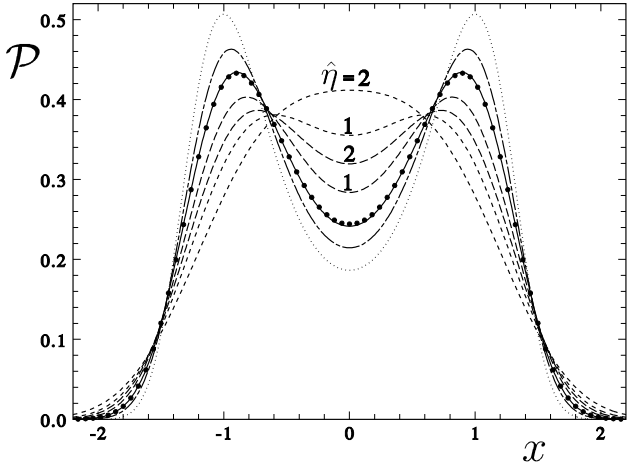,width=82mm,angle=90}}
\caption{Configuration density $\P(x)=\ave{\delta(\hat{x}-x)}$ of the
double-well quartic potential for $g=2$, $t=1$, and different values of
the damping parameters $\hat\eta=1,2$ (as indicated on the picture) and
$\hat\omega_{{}_{\rm{D}}}=1$ (long-dashed lines) and 5 (short-dashed
lines). The solid line is the nondissipative effective potential result
for  $\hat\eta=0$, that agrees with the corresponding exact result
(filled circles); the dotted curve corresponds to the classical limit,
and the dash-dotted curve reports the result of the standard SPR model
from Ref.~\protect\cite{CRTV1997} for an Ohmic dissipative strength
$\hat\gamma\equiv\gamma/(m\omega_0)=5$.
 \label{f.dwrho} }
\end{figure}

In Fig.~\ref{f.dwrho} we report the shape of the coordinate probability
distribution $\P(x)=\ave{\delta(\hat{x}-x)}$, for selected values of
the dissipation strength $\hat\eta$ and bandwidth
$\hat\omega_{{}_{\rm{D}}}$, at the coupling $g=2$; this gives a ground
state energy $e_0=0.69797$, with the next excited level at
$e_1=1.04637$, above the barrier. When dissipation is switched on
$\P(x)$ tends to go farther away from the classical distribution,
$\P_{\rm{c}}\sim{e}^{-v(x)/t}$, due indeed to the enhanced fluctuations
of the coordinate, and the two-peaked structure is eventually lost.
This happens by rising either $\hat\eta$ or $\hat\omega_{{}_{\rm{D}}}$:
note the strong influence of the latter in increasing the fluctuations,
which diverge in the Ohmic limit $\hat\omega_{{}_{\rm{D}}}\to\infty$.
The figure also reports the opposite result for the standard
dissipative model, obtained as in Ref.~\cite{CRTV1997}, which clearly
shows the suppression of the coordinate fluctuations towards the
classical distribution.

%===== Figure 1 =====
%===== End Figure 1 =====

\begin{figure}[hbt]
\centerline{\psfig{bbllx=15mm,bblly=32mm,bburx=186mm,bbury=249mm,%
 figure=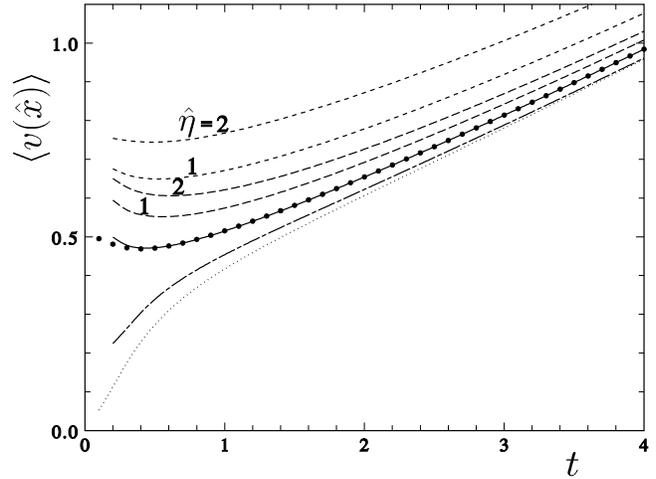,width=82mm,angle=90}}
\caption{Average potential energy $\ave{v(\hat{x})}$ of the double-well
quartic potential vs reduced temperature $t$, for $g=2$ and different
values of $\hat\eta$ and $\hat\omega_{{}_{\rm{D}}}$. Curves and symbols
as in Fig.~\protect\ref{f.dwrho}.
 \label{f.dwavev} }
\end{figure}

Typical results found for the average potential energy
$v(t)=\ave{v(x)}$ are displayed in Fig.~\ref{f.dwavev}. In the
nondissipative case, comparison with the exact data shows that the
effective potential gives very accurate results, in spite of the strong
coupling. The curves for the anomalous dissipative system show an
increase of the potential energy, corresponding to the higher
fluctuations of the coordinate, while for the standard SPR model the
opposite occurs.

%===== Figure 2 ====
%===== End Figure 2 =====

\subsection{$\phi^4$ chain}
\label{ss.phi4}

The quantum $\phi^4$ chain with standard dissipation was studied in
Ref.~\cite{CFTV1999}; the effective potential approach was previously
applied in the nondissipative case~\cite{GTV1988fi4,CGTVV1995} and
checked through quantum Monte Carlo simulations~\cite{JankeS1995}. It
can be viewed as the discretized version of a continuum nonlinear field
theory, described by the (undamped) Lagrangian
\begin{equation}
 {\cal L} = G\,a ~\sum_{i=1}^N \bigg[\frac{\,\dot q_i^2}2
 -\frac{c^2}{2a^2}\,(q_i-q_{i-1})^2 + \frac{\Omega^2}8\, v(q_i) \bigg]~,
\label{e.Lphi4}
\end{equation}
where $a$ is the chain spacing, $c$ is the `relativistic' velocity,
$\Omega$ is the gap of the bare dispersion relation, $G$ is a constant
with the dimension (mass$\,{\times}\,$length), and periodic boundary
conditions ensure translation symmetry. The local nonlinear potential
\begin{equation}
 v(x)= \big(1-x^2\big)^2
\label{e.vphi4}
\end{equation}
has two wells in $x_{\rm{m}}=\pm{1}$ [with $v''(x_{\rm{m}})=8$], so
that the classical $\phi^4$ chain has two degenerate
translation-invariant minimum configurations, $\{q_i=1\}$ and
$\{q_i=-1\}$, as well as relative minima connecting the two wells, the
static `kinks'. In the continuum limit $ia\to{z}$ the kink
configuration is indeed $q(z)=\pm\tanh[\Omega(z-z_0)/c]$, so that it is
localized with a characteristic length $c/\Omega$ and its energy is
$\varepsilon_{{}_{\rm{K}}}=2G\Omega{c}/3$.

The ratio between the characteristic frequency $\Omega$ of the
quasiharmonic excitations of the system and the energy scale
$\varepsilon_{{}_{\rm{K}}}$ defines the quantum coupling parameter
$Q=\Omega/\varepsilon_{{}_{\rm{K}}}=3/(2Gc)$; the discreteness of the
chain is measured by the kink length in lattice units,
$R=c/(\Omega{a})$ ($R\to\infty$ in the continuum limit). Moreover,
$t\equiv{T}/\varepsilon_{{}_{\rm{K}}}$ is the reduced temperature used
from now on. The most interesting features appear when kinks are
excited in the system; for instance, they cause a peak in the classical
specific heat at $t\sim{0.2}$.

Using the dimensionless quantities just introduced, the Weyl symbol for
the undamped $\phi^4$ Hamiltonian can be written as
\begin{eqnarray}
 \frac{\H}{\varepsilon_{{}_{\rm{K}}}}
 &=& \frac{Q^2R}3\sum_{i=1}^N p_i^2 + V(\bq)
\\
 \frac{V(\bq)}{\varepsilon_{{}_{\rm{K}}}}
 &=& \frac3{2R}\sum_{i=1}^N\bigg[
 \frac{R^2}2\,(q_i-q_{i-1})^2 + \frac{v(q_i)}8 \bigg]~,
\label{e.Vphi4}
\end{eqnarray}
where the momenta are such that $[\hat q_i,\hat p_j]=i\,\delta_{ij}$,
and Eq.~(\ref{e.Ak}) gives
$m_k^{-1}=2Q^2R\varepsilon_{{}_{\rm{K}}}/3\equiv{m}^{-1}$.

As for dissipation, $N$ identical independent environmental baths
coupled to each momentum are assumed, giving a diagonal kernel matrix
$\kappa_{n,ij}=\delta_{ij}\,\kappa_n$, so Eq.~(\ref{e.Kk}) gives
$\kappa_{n,k}=\kappa_n$. We take
$\kappa_n=|\nu_n|\,\tilde\eta\big(|\nu_n|\big)$ with the Drude
form~(\ref{e.Drude}) for $\tilde\eta$; the dimensionless quantity
$m_k\kappa_{n,k}=m\kappa_n\equiv{a}_n$ can eventually be written as in
Eq.~(\ref{e.an}), so that dissipation is characterized by the
dimensionless parameters
$\hat\omega_{{}_{\rm{D}}}=\omega_{{}_{\rm{D}}}/\Omega$ and
$\hat\eta=3\eta/(2QR)$.

From Eqs.~(\ref{e.Bij}) and~(\ref{e.Bk}) the LCA renormalized
frequencies are found to be
\begin{equation}
 \omega^2_k=\Omega^2\,[1-3D(t)+4R^2\sin^2(k/2)]~,
\end{equation}
while the relevant renormalization coefficient $D(t)\equiv{C}_{ii}$
follows from Eqs.~(\ref{e.alphakLCA}) and~(\ref{e.CkLCA}),
\begin{equation}
 D(t) = \frac{Q^2R}{3\,t}\frac1N\sum_k\sum_{n=1}^\infty
 \frac{1+a_n}{(\pi n)^2+(1{+}a_n)\,f^2_k} ~,
\label{e.Dphi4}
\end{equation}
where $f_k=\beta\omega_k/2=(Q/2t)\big[1-3D(t)+4R^2\sin^2(k/2)\big]$;
these two self-consistent equations can be solved numerically, and it
appears that the approximation is meaningful when $3D\ll{1}$.
Eventually, the LCA effective potential can be written as the original
one, just replacing the local interaction $v(x)$ by
\begin{equation}
 v_{\rm{eff}}(x) = \big[1-3D(t)-x^2\big]^2 + 6 D^2(t) + t \tilde\mu(t) ~,
\end{equation}
with
\begin{equation}
 \tilde\mu(t) = \frac{16R}3\frac1N \sum_k \sum_{n=1}^{\infty} \ln\,
 \frac{(\pi n)^2+(1{+}a_n)\,f^2(x)}{(\pi n)^2}~.
\label{e.muphi4}
\end{equation}

\begin{figure}
\centerline{\psfig{bbllx=13mm,bblly=30mm,bburx=161mm,bbury=257mm,%
figure=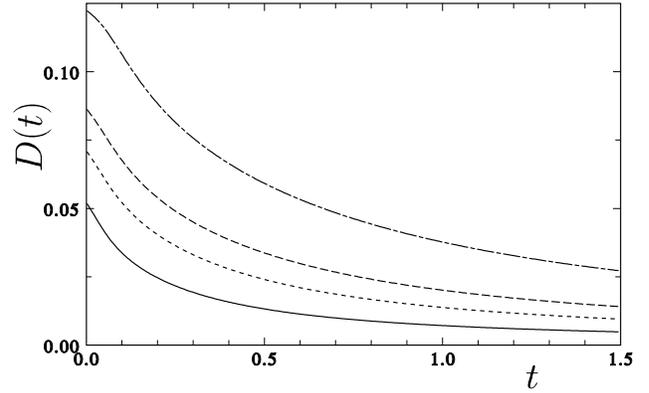,width=82mm,angle=90}}
 \caption{Renormalization coefficient $D(t)$ [Eq.~(\protect\ref{e.Dphi4})]
vs reduced temperature $t$, for different values of the damping
strength $\hat\eta$, at fixed coupling parameter $Q=0.2$, kink length
$R=5$, and Drude cutoff frequency $\hat\omega_{{}_{\rm{D}}}=1$. Solid
line: $\hat\eta=0$ (undamped); short-dashed line: $\hat\eta=1$;
long-dashed line: $\hat\eta=2$; dash-dotted line: $\hat\eta=5$. Note
that $D(t)$ increases when $\hat\eta$ increases.
\label{f.d0} }
\end{figure}

\begin{figure}
\centerline{\psfig{bbllx=12mm,bblly=29mm,bburx=190mm,bbury=260mm,%
figure=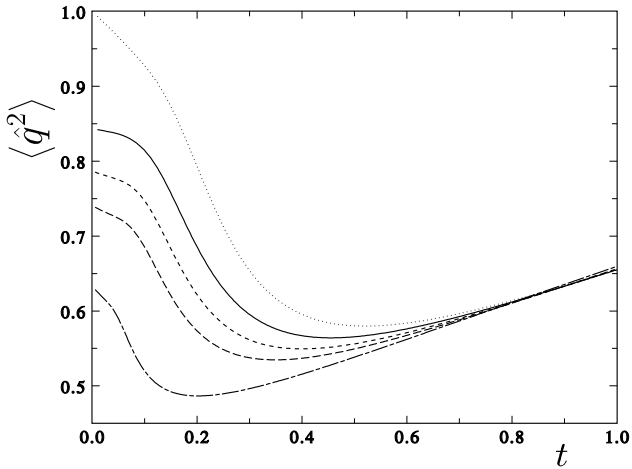,width=82mm,angle=90}}
 \caption{Mean-square fluctuations of the coordinate operator
$\langle\hat{q}_i^2\rangle$ vs reduced temperature $t$. Parameters and
lines as in Fig.~\protect\ref{f.d0}. The dotted line represents the
classical result ($Q{=}0$).
\label{f.q2med} }
\end{figure}

\begin{figure}
\centerline{\psfig{bbllx=15mm,bblly=30mm,bburx=189mm,bbury=256mm,%
figure=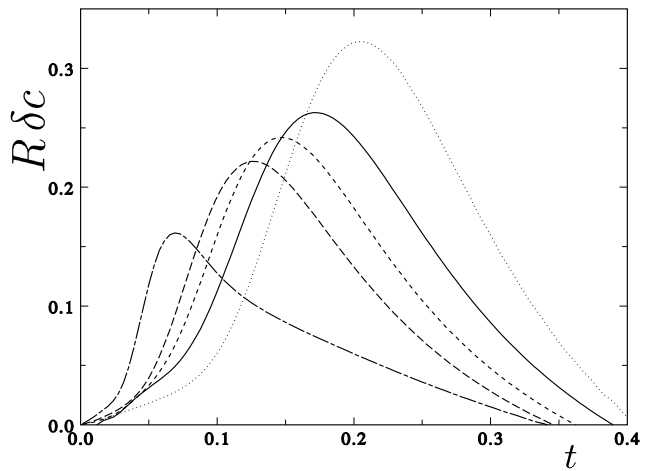,width=82mm,angle=90}}
 \caption{Nonlinear contribution to the specific heat, defined as
$\delta{c}(t)=c(t)-c_{{}_{\rm{H}}}(t)$ (times the kink length $R$), vs
reduced temperature $t$. Parameters and lines as in
Fig.~\protect\ref{f.d0}. The dotted line represents the classical
result ($Q{=}0$). Increasing $\hat\eta$, $\delta{c}(t)$ goes farther
away from the classical behavior, as if $Q$ were
raised~\protect\cite{GTV1988fi4}.
\label{f.cnl} }
\end{figure}

For any parameter set ($t$, $Q$, $R$, $\hat\eta$,
$\hat\omega_{{}_{\rm{D}}}$) the self-consistent computation of $D$ and
hence of the last term of $\tilde\mu$, which takes a negligible
computer time using a continuum termination of the $n$ summation,
completely determines the effective potential $v_{\rm{eff}}(x)$. The
problem is then ready for a numerical evaluation of
Eq.~(\ref{e.mdf.aveO}), that gives the partition function (setting
$\hat\O=1$) and the thermal averages of observables. We employed the
numerical transfer matrix technique~\cite{SchneiderS1980}, that reduces
the configuration integral for a one-dimensional array with
nearest-neighbor interaction to a secular integral equation. The
evaluation of the latter is implemented numerically using a discrete
mesh for the values of each degree of freedom. Temperature scans over
the region of interest were performed and several thermodynamic
quantities were calculated, taking the value `per site' for the
extensive ones. In particular, the internal energy
$u(t)=f(t)-t\partial_tf(t)$ and the specific heat
$c(t)=\partial_tu(t)=-t\partial_t^2f(t)$ follow from the free energy
per site, $f(t)=-N^{-1}\,t\,\ln\Z(t)$ by numerical derivation, while
Eq.~(\ref{e.mdf.aveO}) was used to calculate the thermal average of the
squared site coordinate $\bave{\hat{q}_i^2}$ and of the local potential
$\bave{v(\hat{q}_i)}$, the square nearest-neighbor displacement
$\bave{(\hat{q}_i-\hat{q}_{i-1})^2}$, and the square momentum
$\bave{\hat{p}_i^2}$. The quantities reported in Figs.~\ref{f.d0},
\ref{f.cnl}, and~\ref{f.q2med} are evaluated for fixed values of the
kink length $R=5$ and of the quantum coupling $Q=0.2$, which give
fairly strong quantum effects. For comparison, we also report the
classical result corresponding to $Q=0$. Moreover, a fixed
representative value of the Drude cutoff frequency
$\hat\omega_{{}_{\rm{D}}}=1$ is used, in order to analyze the
dependence upon the dissipation strength $\hat\eta$.

As we already remarked, the pure-quantum fluctuations of the coordinate
are made stronger by the anomalous dissipative coupling. This appears
in Fig.~\ref{f.d0}, where the pure-quantum renormalization coefficient
$D(t)=\bdave{\xi_i^2}$ of Eq.~(\ref{e.Dphi4}) is reported for different
values of $\hat\eta$, starting from the nondissipative value
$\hat\eta=0$. The temperature dependence of the coefficient $D(t)$
affects the effective potential and the free energy, allowing us to
describe how the classical behavior of thermal averages is affected by
both quanticity and dissipation.

%===== Figure 3 ====
%===== End Figure 3 =====

In Fig.~\ref{f.q2med} the temperature behavior of the mean-square
fluctuations of the site coordinate $\langle\hat{q}_i^2\rangle$ is
reported. At $t=0$, in the classical case the coordinate lies in the
minima $\{q^2_i=1\}$ of the potential (\ref{e.Vphi4}), while in the
quantum non-dissipative case the value at $t=0$ is smaller,
$\langle\hat{q}_i^2\rangle=1-3D\simeq{0.84}$: this corresponds to the
minima of $v_{\rm{eff}}(q)$ and reflects the fact that quantum
fluctuations make the configuration to climb the barrier rather than
the steeper walls. Switching on the temperature enhances the same
effect and $\langle\hat{q}_i^2\rangle$ decreases at finite temperature,
until when at $t\gtrsim{0.5}$ many kinks are excited in the system and
the coordinate distribution begins to spread towards the walls, causing
the subsequent increase from a minimum value $\sim{0.56}$ at
$t\sim{0.45}$, eventually collapsing onto the classical curve. At
variance with the case of standard dissipation~\cite{CFTV1999}, when
the damping strength $\hat\eta$ is switched on, a further enhancement
of the distribution spread occurs: this effect could be qualitatively
interpreted as an `effective' increase of the quantum coupling and
allows $\langle\hat{q}_i^2\rangle$ to reach smaller values, at smaller
temperature.

%===== Figure 4 ====
%===== End Figure 4 =====

On the other hand, the fluctuations of the momenta are quenched by
dissipation, and the role of the damping effects is nonpredictable on a
simple basis if one considers thermodynamic quantities where both
coordinates and momenta enter into play. One of these is the specific
heat $c(t)$. Its {\em nonlinear} part, namely its total value minus the
corresponding (dissipative) harmonic contribution,
$\delta{c}(t)=c(t)-c_{\rm{h}}(t)$, is very sensitive to the
nonlinearity of the system, since its value is zero in a harmonic
approximation. The fact that our approach retains all classical
nonlinear features is crucial for getting $\delta{c}$, whose
temperature behavior is reported in Fig.~\ref{f.cnl}. Switching on and
increasing the strength of the environmental coupling $\hat\eta$ the
curves can be seen to go farther apart from the classical one, which is
a nontrivial result. A physical explanation is that anomalous
dissipation increases the uncertainty in the kink positions,
substantially making them longer and effectively raising their mutual
interaction.

%===== Figure 5 ====
%===== End Figure 5 =====

%================================================================
\section{Conclusions}

In this paper we have discussed the problem of {\em anomalous
dissipation}, previously considered by Leggett~\cite{Leggett1984} and
derived the effective potential formalism for an interacting many-body
system with this kind of dissipation, which can arise from a
microscopic description of the system-bath coupling, as done for
instance in Ref.~\cite{SolsZ1997}.

Our approach allows one to reduce quantum mechanical thermodynamic
calculations to a classical-like configuration integral, where both
quantum and dissipative effects are included. In order to deal with
many degrees of freedom the necessary {\em low-coupling} approximation
has been introduced. The latter, if the system's symmetries are
exploited, results in very simple expressions for the renormalization
coefficients appearing in the theory. This is shown in detail for the
case of translation symmetry.

Applications of the framework have been made to the single particle in
double-well potential and to a discrete $\phi^4$ one-dimensional field,
whose strong nonlinearity yields kink excitations that play a dominant
role. In both examples a Drude-like spectrum of the environmental
coupling has been chosen for dissipation, and its influence on thermal
quantities has been analyzed.

The effective potential is a unique tool for dealing with such a
system, since any approximate theory must retain the strong
nonlinearity, which has a mainly classical character, and this rules
out conventional perturbative approaches. The method proved to be
useful for the above model systems, so it is expected that it will find
application in up-to-date physical problems. A possible application of
the formalism regards the effect of the blackbody electromagnetic
field~\cite{SolsZ1997} onto the phase diagram of two-dimensional
Josephson junction arrays.

\acknowledgements
 This research was supported by the COFIN2000-MURST fund.

%================================================================
\appendix

\section{Standard SPR model}
\label{a.sSPR}

Eliminating the bath variables from the equations of motion for the
standard SPR model~\eqref{e.sSPR} one gets~\cite{FordLO1988} the
Langevin equation~\eqref{e.sLangevin}, where the coupling with the bath
is described by the memory function
$\gamma(t)=\Theta(t)\sum_\ell{m_\ell}\omega_\ell^2\cos(\omega_\ell\,t)$,
whose Laplace transform is
\begin{equation}
 \tilde\gamma(z)=
 \sum_\ell m_\ell \omega_\ell^2~ \frac z{z^2+\omega_\ell^2}~,
\label{e.stildegamma}
\end{equation}
and by the Gaussian random force $\hat{F}(t)$, whose time correlators
can be expressed in terms of the memory function by spectral
relations~\cite{FordLO1988} depending on the bath's temperature.

The quantum thermodynamics, after the bath has been `traced out' in the
path-integral for~(\ref{e.sSPR}), is described by the density
matrix~\cite{Weiss1999}
\begin{eqnarray}
 \rho(q'',q')&=&\int_{q'}^{q''}\!\! \D [q(u)]
 ~e^{-S[q(u)]-S_{{}_{\rm{I}}}[q(u)]}
\label{e.srho}
\\
 S[q(u)] &=&
 \int_0^{\beta\hbar}\!\frac{du}{\hbar}\,
 \Big[\frac m2\dot q^2(u)+V\big(q(u)\big)\Big]
\end{eqnarray}
and $S_{{}_{\rm{I}}}$ given in Eq.~\eqref{e.sinfluence}; the Matsubara
transform of the kernel $k(u)$ reads
\begin{equation}
 k_n = \int_0^{\beta\hbar} \frac{du}{\beta\hbar} ~k(u)\,e^{-i\nu_nu}
  = \sum_\ell m_\ell \omega_\ell^2~ \frac
 {\nu_n^2}{\nu_n^2+\omega_\ell^2}
\label{e.skn}
\end{equation}
and is thus given by $k_n=|\nu_n|~\tilde\gamma\big(|\nu_n|\big)$. For a
harmonic potential, $V(q)=m\omega^2\,q^2/2$, the evaluation of the path
integral can be performed exactly and the density matrix turns out to
be a Gaussian with partition function and variances as
follows~\cite{Weiss1999}:
\begin{eqnarray}
 \Z_{{}_{\rm{H}}} &=& \frac1{\beta\hbar\omega}~
 \prod_{n=1}^\infty \frac {\nu_n^2}{\nu_n^2+\omega^2+k_n/m} ~,
\nonumber\\
 \ave{p^2}_{{}_{\rm{H}}} &=& \frac m\beta \sum_{n=-\infty}^\infty
 \frac {\omega^2+k_n/m}{\nu_n^2+\omega^2+k_n/m} ~,
\nonumber\\
 \ave{q^2}_{{}_{\rm{H}}} &=& \frac 1{\beta m} \sum_{n=-\infty}^\infty
 \frac 1{\nu_n^2+\omega^2+k_n/m} ~.
\label{e.sSPR-HA}
\end{eqnarray}
In the nondissipative limit, $k_n{\to}0$, the summations give the
well-known results $\Z=1/(2\sinh{f})$,
$\ave{\hat{q}^2}=\frac\hbar{2m\omega}\,\coth{f}$ and
$\ave{\hat{p}^2}=\frac{\hbar{m}\omega}2\,\coth{f}$, with
$f=\beta\hbar\omega/2$.

\section{Anomalous SPR model}
\label{a.aSPR}

From (\ref{e.aSPR}) we get the equations of motion
\begin{eqnarray}
 ~m\,\ddot{\hat q} + V'(\hat q) &=&
  \sum_\ell \frac m{m_\ell}
  \big[ m_\ell\omega_\ell^2\,\hat q_\ell-V'(\hat q) \big] ~,
  \label{e.ddotq}
\\
 ~m_\ell \big(\ddot{\hat q}_\ell + \omega_\ell^2\,\hat q_\ell\big) &=&
  V'(\hat q) ~;
  \label{e.ddotqell}
\end{eqnarray}
the second one has the general solution
$\hat{q}_\ell(t)=\hat{q}_\ell^{\rm{h}}(t)+\hat{q}_\ell^{\rm{p}}(t)$,
with the retarded form of the particular solution,
\begin{eqnarray}
 m_\ell\omega_\ell^2~\hat q_\ell^{\rm{p}}(t) &=&
 V'\big(\hat q(t)\big)
\nonumber\\
 &-& \int_{-\infty}^t \!\! dt'\,\cos\big[\omega_\ell(t{-}t')\big]
 ~\partial_{t'} V'\big(\hat q(t')\big) ~,
\end{eqnarray}
and the homogeneous solution
\begin{equation}
 \hat q_\ell^{\rm{h}}(t)= \hat q_\ell~\cos (\omega_\ell t)
 + \frac {\hat p_\ell}{m_\ell\omega_\ell}~\sin(\omega_\ell t)~.
\end{equation}
Inserting this result in Eq.~(\ref{e.ddotq}) one gets
Eq.~\eqref{e.aLangevin}, where
\begin{equation}
 \eta(t) =
 \Theta(t) \sum_\ell \frac 1{m_\ell}\, \cos(\omega_\ell t)~,
\label{e.eta}
\end{equation}
while the fluctuating force term,
\begin{equation}
 \hat F(t)= m \sum_\ell \omega_\ell^2~\hat q_\ell^{\rm{h}}(t)
\end{equation}
is a Gaussian random process, whose correlations arise from the
assumption that the initial values $\{\hat{q}_\ell,\hat{p}_\ell\}$
correspond to the equilibrium at the temperature $T=\beta^{-1}$ for the
isolated bath, namely
$\ave{\hat{q}_\ell\hat{q}_{\ell'}}=\delta_{\ell\ell'}
~\frac\hbar{2m_\ell\omega_\ell}\,\coth\frac{\beta\hbar\omega_\ell}2$~,
$\ave{\hat{p}_\ell\hat{p}_{\ell'}}=\delta_{\ell\ell'}
~\frac{\hbar{m}_\ell\omega_\ell}2\,\coth\frac{\beta\hbar\omega_\ell}2$~,
$\ave{\hat q_\ell\hat p_{\ell'}}=\delta_{\ell\ell'}~\frac{i\,\hbar}2$~.

The thermal density matrix for the harmonic oscillator,
$V(q)=m\omega^2\,q^2/2$, can be evaluated reducing the problem to the
standard-dissipation one, just by performing the canonical
transformation $\hat{q}\to-\hat{p}/(m\omega)$ and
$\hat{p}\to{m}\omega\,\hat{q}$; to get exactly the same form of
Eq.~(\ref{e.sSPR}) it is sufficient to replace
$\hat{p}_\ell\to{m}\omega\,\hat{q}_\ell$,
$\hat{q}_\ell\to-\hat{p}_\ell/(m\omega)$, and
$m_\ell\to(m\omega)^2/(m_\ell\omega_\ell^2)$. One finds, of course,
Eqs.~(\ref{e.sSPR-HA}) with the exchange of the expressions
$\ave{q^2}_{{}_{\rm{H}}}\to(m\omega)^{-2}\ave{p^2}_{{}_{\rm{H}}}$ and
$\ave{p^2}_{{}_{\rm{H}}}\to(m\omega)^{2}\ave{q^2}_{{}_{\rm{H}}}$. Due
to the above redefinition of the bath masses $\{m_\ell\}$, the
kernel~(\ref{e.skn}) takes the form $k_n=(m\omega)^2\,\kappa_n$, with
\begin{equation}
 \kappa_n = \sum_\ell \frac 1{m_\ell}
 ~\frac {\nu_n^2}{\nu_n^2+\omega_\ell^2} ~,
\label{e.akn}
\end{equation}
so eventually one has
\begin{eqnarray}
 \Z_{{}_{\rm{H}}} &=& \frac1{\beta\hbar\omega}~
 \prod_{n=1}^\infty \frac {\nu_n^2}{\nu_n^2+(1+m \kappa_n)\,\omega^2} ~,
 \label{e.aZH}
\\
 \ave{p^2}_{{}_{\rm{H}}} &=& % \lambda_{{}_{\rm{H}}} =
 \frac {m}\beta \sum_{n=-\infty}^\infty
 \frac {\omega^2}{\nu_n^2+(1+m \kappa_n)\,\omega^2} ~,
\label{e.alambdaT}
\\
 \ave{q^2}_{{}_{\rm{H}}} &=& % \alpha_{{}_{\rm{H}}} =
 \frac 1{m\beta} \sum_{n=-\infty}^\infty
 \frac {1+m \kappa_n}{\nu_n^2+(1+m \kappa_n)\,\omega^2} ~.
\label{e.aalphaT}
\end{eqnarray}

\section{Anomalous influence action}
\label{a.SI}

For the bath part of Eq.~(\ref{e.aSPR}) the Feynman-Kac expression in
terms of the momentum can be used: the definition of the influence
action is therefore
\begin{equation}
 e^{-S_{{}_{\rm{I}}}[p(u)]}=\frac 1{\Z_{{}_{\rm{B}}}}
 \prod_\ell\oint \D [p_\ell(u)]
 ~e^{-S_{{}_{\rm{B}}}\big[\{p_\ell(u)\}\big]}~,
\end{equation}
where $\Z_{{}_{\rm{B}}}=\prod_\ell(2\sinh\omega_\ell)^{-1}$ is the
partition function of the isolated bath, and the bath action reads
\begin{equation}
 S_{{}_{\rm{B}}} = \int_0^{\beta\hbar} \frac{du}\hbar~\sum_\ell
 \frac 1{2m_\ell\omega_\ell^2}
  \Big[\dot p_\ell^2(u) +
  \omega_\ell^2 \big(p_\ell(u)-p(u)\big)^2 \Big] ~.
\end{equation}
Performing the Fourier transformation into Matsubara components
$p(u)=\sum_n p_n~e^{i\nu_n\,u}$ [and same for $p_\ell(u)$] one gets
\begin{eqnarray}
 S_{{}_{\rm{B}}} &=& \sum_\ell \frac \beta{2m_\ell\omega_\ell^2}
 \sum_n
 \Big[\nu_n^2\,|p_{\ell n}|^2 + \omega_\ell^2 |p_{\ell n}-p_n|^2 \Big]
\\
 &=& \sum_\ell \frac \beta{2m_\ell\omega_\ell^2}
 \sum_n
 \bigg[(\nu_n^2+\omega_\ell^2)\,|\tilde p_{\ell n}|^2
 + \frac {\nu_n^2\omega_\ell^2}
 {\nu_n^2+\omega_\ell^2}\,|p_n|^2 \bigg] ,
\nonumber
\end{eqnarray}
where the variable shift $\tilde{p}_{\ell{n}}=p_{\ell{n}}
-\frac{\omega_\ell^2}{\nu_n^2+\omega_\ell^2}\,p_n$ has been performed.
The path integral over $\tilde{p}_\ell(u)$ gives just the partition
function $\Z_{{}_{\rm{B}}}$, so one is left with the {\em anomalous}
influence action
\begin{equation}
 S_{{}_{\rm{I}}}[p(u)] = \frac \beta 2
 \sum_{n=-\infty}^\infty ~\kappa_n~|p_n|^2~,
\end{equation}
i.e., Eq.~\eqref{e.aSI}, where $\kappa_n$ is again given by
Eq.~\eqref{e.akn}, so that comparing with the Laplace transform of the
memory function~(\ref{e.eta}),
\begin{equation}
 \tilde\eta(z)= \sum_\ell \frac 1{m_\ell}
 ~ \frac z{z^2+\omega_\ell^2}~,
\label{e.atildeeta}
\end{equation}
one has $\kappa_n=|\nu_n|~\tilde\eta\big(|\nu_n|\big)$~, i.e.
Eq.~\eqref{e.akneta}.

\section{Evaluation of $\bar\rho_0$ for one degree of freedom}
\label{a.1df.rho0p}

Let us start from the definition (\ref{e.barrho}) of the reduced
density in terms of the system's trial action, given by
Eq.~(\ref{e.Spq}) with the trial potential~(\ref{e.V0}),
\begin{eqnarray}
 \hbar S_0 &=& \!\!\int_0^{\beta\hbar}\!\!\!du
 \Big[\, {\textstyle \frac i2} \big(q_u\dot p_u{-}p_u\dot q_u\big)
 +\frac {p_u^2}{2m}+w+\frac{m\omega^2(q_u{-}\bar{q})^2}2 \Big]
\nonumber\\
 & & \hspace{-3mm}
 +{\textstyle \frac i2}( q_0p_\beta {-} p_0q_\beta )
 +i\big[ (q_\beta{-}q_0)\,p - (p_\beta{-}p_0)\,q \big] ~,
\end{eqnarray}
and of the dissipative action~(\ref{e.aSI}). In the calculations
$\bar{q}$ is to be regarded as a fixed parameter and we can omit to
explicit the dependence of $w$ and $\omega^2$ on it. Performing the
shift $q_u\to\bar{q}+q_u$ and using the Fourier representation
\begin{equation}
 {\textstyle \delta\Big(\int_0^{\beta\hbar}\!
 \frac{du}{\beta\hbar}\,q_u \Big) }
 = \beta m\omega^2 \int \frac{dz}{2\pi}
 ~\exp \Big( {\textstyle \int_0^{\beta\hbar} \!\frac{du}\hbar
 ~i\,m\omega^2z\,q_u } \Big)
\end{equation}
one gets
\begin{eqnarray}
 \bar\rho_0(p,q;\bar{q}) &=& \beta m\omega^2\, e^{-\beta w}
 \int \frac{dz}{2\pi}
\nonumber\\
 & & \hspace{-5mm}
 \times\int \D [p_u,q_u]
 ~ e^{-S_1[p_u,q_u;\bar{q}]-S_{{}_{\rm{I}}}[p_u]}~,
\end{eqnarray}
\begin{eqnarray}
 \hbar S_1 &=& \int_0^{\beta\hbar}\hspace{-3mm} du
 \Big[\frac i2 \big(q_u\dot p_u{-}p_u\dot q_u\big)
 {+}\frac {p_u^2}{2m}{+}\frac{m\omega^2q_u^2}2{-}im\omega^2zq_u\Big]
\nonumber\\
 & & \hspace{-3mm}
 +\frac i2( q_0p_\beta {-} p_0q_\beta )
 +i\big[ (q_\beta{-}q_0)\,p - (p_\beta{-}p_0)\,\xi \big] ~.
\end{eqnarray}
where $\xi\equiv{q}-\bar{q}$. To eliminate the linear term we shift now
$q_u\to q_u+iz$,
\begin{eqnarray}
 \bar\rho_0(p,q;\bar{q}) &=& \beta m\omega^2\, e^{-\beta w}
 \int \frac{dz}{2\pi}~ e^{-\beta m\omega^2 z^2/2}
\nonumber\\
 & & \hspace{-5mm}
 \times \int \D [p_u,q_u]
 ~ e^{-S_2[p_u,q_u;\bar{q}]-S_{{}_{\rm{I}}}[p_u]}~,
\label{e.barrho2}
\end{eqnarray}
so that $S_2$ is the harmonic-oscillator action,
\begin{eqnarray}
 \hbar S_2 &=& \int_0^{\beta\hbar}du\,
 \Big[\, \frac i2 \big(q_u\dot p_u{-}p_u\dot q_u\big)
 +\frac {p_u^2}{2m}+\frac{m\omega^2q_u^2}2\Big]
\nonumber\\
 & & \hspace{-8mm}
 +\frac i2( q_0p_\beta {-} p_0q_\beta )
 {+}i\big[ (q_\beta{-}q_0)\,p -(p_\beta{-}p_0)\,(\xi{-}iz) \big] .
\label{e.S2}
\end{eqnarray}
Now, the point is to get rid of the path integral appearing in
Eq.~(\ref{e.barrho2}), that represents the Weyl symbol for the density
operator of the harmonic oscillator with the dissipative action
$S_{{}_{\rm{I}}}[p_u]$. However, the result for the harmonic oscillator
with anomalous dissipation is known from Eqs.~(\ref{e.aZH}),
(\ref{e.aalphaT}) and~(\ref{e.alambdaT}):
\begin{equation}
 \rho_{{}_{\rm{H}}}(p,q) = \frac{2\pi\,e^{-\mu}}{\beta\omega}
 ~\frac {e^{-p^2/2\lambda_{{}_{\rm{T}}}}}
 {\sqrt{2\pi\lambda_{{}_{\rm{T}}}}}
 ~\frac {e^{-q^2/2\alpha_{{}_{\rm{T}}}}}
 {\sqrt{2\pi\alpha_{{}_{\rm{T}}}}}~,
\end{equation}
where the partition function from~(\ref{e.aZH}) is written as
$\Z=e^{-\mu}/(\beta\hbar\omega)$ and $\mu$ given as in
Eq.~(\ref{e.mu}). One gets therefore
\begin{equation}
 \bar\rho_0(p,q;\bar{q}) = \beta m \omega^2\, e^{-\beta w}
 \int \frac{dz}{2\pi}~ e^{-\beta m\omega^2z^2/2}
 ~\rho_{{}_{\rm{H}}}(p,\xi{-}iz) \,.
\label{e.barrho3}
\end{equation}
This last Gaussian convolution affects the coordinate part and results
in a Gaussian for $\xi$ with the pure-quantum variance
$\alpha=\alpha_{{}_{\rm{T}}}-1/(m\omega^2\beta)$ reported in
Eq.~(\ref{e.alpha}): what is subtracted is just the classical
contribution, i.e., the $n{=}0$ term in Eq.~(\ref{e.aalphaT}). The
final result is just Eq.~(\ref{e.barrho1}).

%================================================================
\section{Evaluation of $\bar\rho_0$ for many degrees of freedom}
\label{a.mdf.rho0p}

We evaluate here the path integral~(\ref{e.mdf.brho}) with the
trial action $S_0$ [i.e., Eq.(\ref{e.mdf.S}) with the
`potential'~(\ref{e.mdf.V0})]; the fixed argument of $w(\bar\bq)$
and of the matrix $\bB(\bar\bq)$ is omitted in the following.
First, the delta function is represented as
\begin{equation}
 {\textstyle
 \delta\big( \bar\bq{-}\int_0^{\beta}\!\frac{du}\beta\,\bq_u \big)}
 = \det(\beta\bB^2)\!\int \!\frac{d\bz}{(2\pi)^N}
 ~e^{i\int_0^\beta du~{}^{\rm{t}}\!\bz\,\bB^2(\bq_u{-}\bar\bq)} ~,
\end{equation}
and the exponential can be incorporated into the action $S_0$; then,
performing the shift $\bq_u\to\bq_u+\bar\bq+i\bz$ and introducing the
fluctuation variable $\bxi=\bq-\bar\bq$ one gets
\begin{equation}
 \bar\rho_0(\bp,\bq;\bar\bq) =
 \frac {e^{-\beta\,w}}{\det\bC_{{}_{\rm{C}}}}
 \int \frac{d\bz}{(2\pi)^N}
 \,e^{-\frac12{}^{\rm{t}}\!\bz\,\bC_{{}_{\rm{C}}}^{-1}\bz}
 \rho_{{}_{\rm{HA}}}(\bp,\bxi{-}i\bz) ~,
\label{e.a.brho1}
\end{equation} where $\bC_{{}_{\rm{C}}}^{-1}\equiv\beta\bB^2$ and
\begin{equation}
 \rho_{{}_{\rm{HA}}}(\bp,\bq)= \! \int\!\D[\bp_u,\bq_u]
 ~ e^{-\Sm[\bp_u,\bq_u;\bp,\bq]
 -S_1[\bp_u,\bq_u;\bar\bq]-S_{{}_{\rm{I}}}[\bp_u] },
\label{e.a.rhoHA}
\end{equation}
with
\begin{equation}
 S_1 = \int_0^{\beta}\!\frac{du}2\,
 \Big[\, {}^{\rm{t}}\!\bp_u \bA^2 \bp_u +
 {}^{\rm{t}}\bq_u \bB^2 \bq_u \Big] ~.
\end{equation}
The path integral~(\ref{e.a.rhoHA}) corresponds to the density
matrix of a harmonic system with anomalous dissipation,
$\rho_{{}_{\rm{HA}}}(\bp,\bq)$. The canonical exchange
$\bq_u\to\bp_u$ and $\bp_u\to-\bq_u$ has the effect of turning the
dissipation into standard,
$S_{{}_{\rm{I}}}[\bp_u]\to{S}_{{}_{\rm{I}}}[\bq_u]$; performing
the same transformation onto the external variables $(\bp,\bq)$
and exchanging the matrices $\bA^2\leftrightarrow\bB^2$, an exact
correspondence with the standard dissipative harmonic system is
established,
$\rho_{{}_{\rm{HA}}}(\bp,\bq)=\rho_{{}_{\rm{HS}}}(\bq,-\bp)$, and
the known result for a harmonic system with standard
dissipation~\cite{CFTV1999}) can be used, eventually getting
\begin{equation}
 \rho_{{}_{\rm{HA}}}(\bp,\bq) =
 \frac{e^{-\mu_{{}_{\rm{H}}}}}{\det(\beta\bA\bB)}
 \frac{e^{-\frac12\,{}^{\rm{t}}\!\bq \bC_{{}_{\rm{H}}}^{-1} \bq}}
 {\sqrt{(2\pi)^N\det\bC_{{}_{\rm{H}}}}}
 \frac{e^{\frac12\,{}^{\rm{t}}\!\bp \bLambda_{{}_{\rm{H}}}^{-1} \bp}}
 {\sqrt{(2\pi)^N\det\bLambda_{{}_{\rm{H}}}}} ,
\end{equation}
where
\begin{eqnarray}
 \mu_{{}_{\rm{H}}} &=& \sum\limits_{n=1}^{\infty}\,
 \ln\,\frac{\det(\nu_n^2+\bPsi_n)}{\nu_n^{2N}}~,
\\
 \bC_{{}_{\rm{H}}} &=& \frac1\beta\sum_{n=-\infty}^\infty
 \bB^{-1}\frac {\bPsi_n}{\nu_n^2+\bPsi_n}\,\bB^{-1} ~,
\\
 \bLambda_{{}_{\rm{H}}} &=&  \frac1\beta\,\sum_{n=-\infty}^\infty
 \bB\,\frac1{\nu_n^2+\bPsi_n}\,\bB ~,
\end{eqnarray}
and
\begin{equation}
 \bPsi_n = \bB\,\big[\bA^2 + \bkappa_n\big]\,\bB ~.
\end{equation}
Eventually, the Gaussian convolution in Eq.~(\ref{e.a.brho1}) washes
out the $n{=}0$ component of $\bC$, leaving
\begin{eqnarray}
 \bar\rho_0(\bp,\bq;\bar\bq) &=&
 \bigg(\frac 1{2\pi\beta}\bigg)^{\frac N2} \frac1{\det\bA}
 ~e^{-\beta[w(\bar\bq)+\mu(\bar\bq)]}
\nonumber\\
 & & \hspace{2mm} \times
 \,\frac{e^{-\frac12\,{}^{\rm{t}}\!\bq \bC^{-1} \bq}}
 {\sqrt{(2\pi)^N\det\bC}}
 \,\frac{e^{\frac12\,{}^{\rm{t}}\!\bp \bLambda^{-1} \bp}}
 {\sqrt{(2\pi)^N\det\bLambda}} ~,
\label{e.mdf.brho0}
\end{eqnarray}
where $\bC=\bC_{{}_{\rm{H}}}-\bC_{{}_{\rm{C}}}$ and
$\bLambda=\bLambda_{{}_{\rm{H}}}$.

%================================================================

\end{document}